# DEVELOPMENT OF INTERACTIVE INSTRUCTIONAL MODEL USING AUGMENTED REALITY BASED ON EDUTAINMENT TO ENHANCE EMOTIONAL QUOTIENT

Nuttakan Pakprod[1] and Panita Wannapiroon[2]

[1] Ph.D student, Information and Communication Technology
for Education Division, Faculty of Technical Education,
King Mongkutt's University of Technology North Bangkok, Thailand
`p_nut8888@homail.com`

[2] Assistant Professor, Information and Communication Technology
for Education Division, Faculty of Technical Education,
King Mongkutt's University of Technology North Bangkok, Thailand
`panita.w@hotmail.com`

## ABSTRACT

*The research aims to develop an interactive instructional model using augmented reality based on edutainment to enhance emotional quotient and evaluate the model. Two phases of the research will be carried out: a development and an evaluation of the model. Samples are experts in the field of IT, child psychology, and 7th grade curriculum management. Ten experts are selected by purposive sampling method. The obtained data are analyzed using mean and standard deviation.*

*The research result demonstrates the following findings:*

*1) The results of this research show that Model consists of 3 elements: IIAR, EduLA, and EQ. EQ is a means to assess EQ based on Time Series Experimental Design using 2 kinds of tools; i.e. EQ Assessment by programs in tablet computers, and EQ Assessment by behavioral observation.*

*(2The ten experts have evaluated the model and commented that the developed model showed high suitability.*

## 1. INTRODUCTION

Education is an important foundation to manage human resource; so the use of technology in education is vital to develop the country. Especially, there should be development of quality of school-age children so that they could have enough professional knowledge, skills, and intelligence to extend their knowledge by themselves. The encouragement for children to adjust themselves in accordance with information under the contexts of changing technology will lead to the lifelong learning system. In addition, they are able to enter the universal education system under amusing learning environment, which will create sustainable, efficient, and enjoyable learning. To prepare the learners for the aforementioned learning, instructors, media, and environment have an important role in the proper interaction with learners, which is the principle of Interactive Instruction. Interactive Instruction [16] and [17] uses Interactive Learning to manage instructional process based on the theories of Constructivism. There are interaction and scaffolding in the activity between instructors and learners, and among the learners themselves. The essence of interactive instruction is the learners will have relation with other group members, who participate in the learning environment set up properly via technology by the instructors. This results in influences on learning, verbal behavior and non-verbal behavior; whereby the said influences may come from stimulation by questions, suggestions, or compliments. As [1] said, interactive learning through playing was significant to promote learning experiences of learners. This is corresponding to [14] who believed that technology would produce efficient interactive instruction for learners.

Augmented Reality [4] is a technology that plays a vital role to support interactive instruction by using virtual reality and vision technology to produce virtuality for learners. This will enables the learners to have new experience, excitement, and thrill in combining digital objects with environments of the real world. Regarding the principles of work, i.e. marking up, searching in Marker Database, processing, and displaying virtual images on different technology equipment, the production of augmented reality requires 4 major parts: markup, sensor scan, processer, and monitor. Markup, or marker, is in the form of graphic such as images, photos, or paintings; whereas sensor scan is a tool to input the prepared data or markup. The said equipment includes various kinds of scanners but nowadays such electronic appliances as webcam, mobile phone, and tablet computer are more popular. Processer, or software, is used to save the markup and data into the database, to compare and process the said markup, and to create virtual images that will be displayed or presented on the monitor. Monitor is a tool that presents data or virtual images already processed by the processer. The said images may be graphic images, 3D images, or motion pictures. This is in line with the research of [6], [9] and [15], who found that augmented reality promoted instruction in science subjects, providing learners with new experiences, excitement, and thrill by mixing objects in digital world with the environments in real life. At present, augmented reality can display anything on a variety of electronic appliances, e.g. Personal computer, notebook, mobile phone, and tablet computer.

Monitor of the said augmented reality is considered the main technology to create interaction among learners, instructors, and contents presented through instructional activity in the suitable environment. The suitable technology [8] for the said activity is a tablet computer because its features are quite similar to those of a personal computer. However, the tablet computer is light and small with a touch screen command. It is also attractive and it can display various formats of information, which will promote self-learning. The main characteristics of tablet computer include: 1) Advanced- both appearance and the data of a tablet computer must be modern and able to get updated all the time, 2) Mobile- it is light and small; so it can be used anytime and anywhere, 3) Electronic note- it can save messages, voices, images, and motion pictures, 4) Personalize- it can save personal information of users and enable them to work with more unity, 5) Less time- it helps users work more within shorter period of time, 6) Protect- it protects users in terms of usage and storage of their information, and 7) Easily- it is easy to use and command through touch screen system and it has only one Home Icon.

The instructional augmented reality with tablet computer offers the learners with entertainment while learning. This is in compliance with the concept of edutainment that promotes learning through amusement with the aid of media and activity. The said instruction must have enthusiastic learning feature and collaborative learning, focus on learning control, and provide the learners with real environment and experiences. The activity must generate high satisfaction, focus on procedures rather than targets, control itself, promote imagination, enable learners to take action, and can be adapted. The learners must take part in decision making, activity, interest, and assessment. The main characteristics of edutainment include: 1) Knowledge- information and presentation must be appropriate and comply to the objectives, 2) Enjoy with Environment- the environment should be set up with an aim to let the learners have amusement, 3) Entertainment- the activity must be relaxing and entertaining, enable collaboration among learners, and focus on what the learners will get rather than the results of activity, and 4) Learners- the learners are free to do the activity that suits their ability, with no force nor pressure. This is in compliance with the study of [5], [7], [10], and [18], who found that the children can naturally learn well through playing. This is because they were happy and ready to receive useful information, which enabled them to improve their learning more effectively.

The outstanding characteristics of interactive instruction, augmented reality, and edutainment will enhance the learners' emotional quotient, the ability to live with others creatively and happily. This corresponds to the study of [2], [11], [12] and [13], who found that emotional quotient helped the learners improve their basic skills, better realize their need, promote their proper expression, interact well with others, do the right things rather than the preferred ones, and look at the brighter side.

The above paragraphs manifest the necessity of a development of an interactive instructional model using augmented reality based on edutainment to enhance emotional quotient. It involves instructional model research and development including conceptualization, model formulation, and presentation of the model testing result to ensure that the developed model is valid for effective use.

## 2. OBJECTIVES

2.1 To develop an interactive instructional model using augmented reality based on edutainment to enhance emotional quotient.

2.2 To evaluate the developed interactive instructional model using augmented reality based on edutainment to enhance emotional quotient.

## 3. RESEARCH SCOPE

3.1 Population is experts in the field of information technology, edutainment, child psychology, and $7^{th}$ grade curriculum management.

3.2 Samples are ten (10) experts in the field of information technology, edutainment, child psychology, and $7^{th}$ grade curriculum management, selected by Purposive Sampling method. The selection criteria are (1) at least Ph.D. degree holder and (2) at least 3 years of relevant experience.

3.3 Variables
Independent variable is the interactive instructional model using augmented reality based on edutainment to enhance emotional quotient.
Dependent variable is the evaluation of the model.

## 4. RESEARCH METHODOLOGY

The instructional model development is carried out in two phases.
Phase 1: The development of an interactive instructional model using augmented reality based on edutainment to enhance emotional quotient.

1.1 Relevant papers and research works are studied, analyzed, and synthesized to formulate a concept of the model development.

1.2 A model is developed based on the data obtained from the research study used in the formulation of the model development concept.

1.3 The model is presented to thesis advisor for consideration and modified it as guided.

1.4 A tool is built to evaluate the model's suitability.

Phase 2: Evaluation and certification of the developed interactive instructional model using augmented reality based on edutainment to enhance emotional quotient.

2.1 The model is submitted to the experts for review and evaluate the suitability.

2.2 The model is modified according to the experts' suggestions.

2.3 After modification, the model is presented in form of narrative diagram.

2.4 Evaluation result on suitability is analyzed using mean ($\bar{X}$) and standard deviation (S.D.). Five Likert-type levels of measurement are identified to assess the model's suitability, namely Strongly agree, Agree, Neither agree nor disagree, Disagree, and Strongly disagree.

## 5. CONCLUSION

The research results are presented in two parts.

Part 1 Interactive Instructional Model Using Augmented Reality based on Edutainment to Enhance Emotional Quotient

The instructional model consists of 3 main elements: (1Interactive Instructional Augmented Reality, (2Edutainment Learning Activity, and (3Emotional Quotient; as shown in Figure1

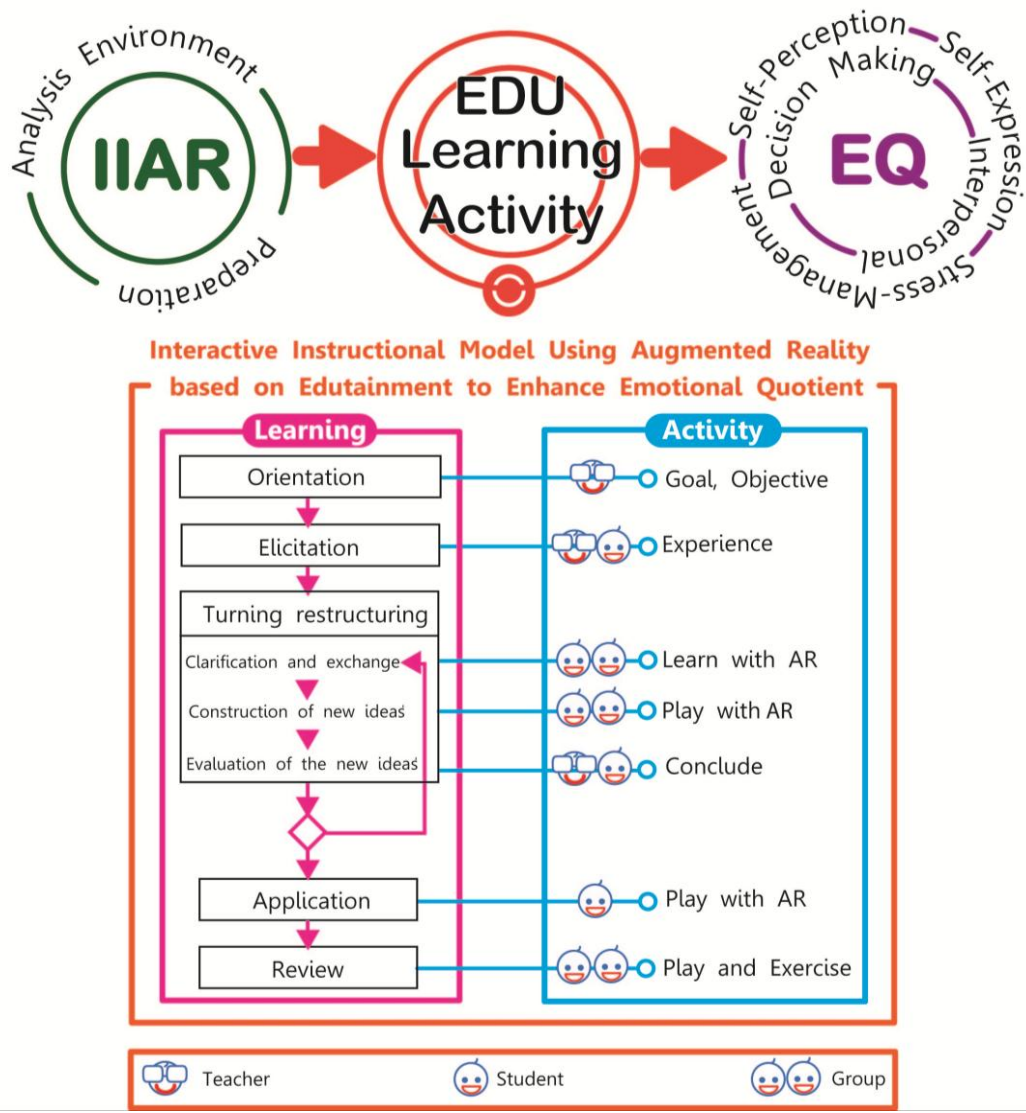

Figure 1 illustrates the edutainment instructional model using learning object
for electronic book on tablet computer to develop emotional quotient.

1. Interactive Instructional Augmented Reality includes: (1Analysis Environment, and (2 Preparation.

1.1 Analysis Environment consists of Needs Analysis, Context Analysis, Learner Analysis, Content Analysis, and Task Analysis.

1.2 Preparation refers to the establishment of learning objectives, contents, and design of instructional management plan based on suitable environment in order to generate the interactive instruction using augmented reality based on edutainment to enhance emotional quotient.

.2Edutainment Learning Activity includes: (1Orientation, (2Elicitation of the prior knowledge, (3Turning restructuring of ideas, (4Application of ideas, and (5Review.

2.1Orientation introduces the learners to the learning environment that is amusing, informal, and relaxing.

2.2Elicitation of the prior knowledge helps review the learners' knowledge or existing experiences that are relevant to the contents and subjects to learn.

2.3Turning restructuring of ideas is a step in which the learners change or increase their knowledge. This step includes clarification and exchange of ideas, construction of new ideas, and evaluation of the new ideas.

2.4 Application of ideas refers to the ability to apply the knowledge in new experiences with happiness.

2.5 Review refers to the increase of knowledge according to the experiences of learners.

.3Emotional Quotient is the evaluation of EQ using 2 kinds of tools; i.e. Emotional Quotient Assessment by programs in tablet computers, in which the learners evaluate themselves, and Emotional Quotient Assessment by behavioral observation, in which the instructors evaluate the learners by EQ behavioral observation form. The Emotional Quotient of learners are assessed in terms of (1Self-Perception (2Self-Expression (3Stress-Management (4Decision Making and (5Interpersonal.

3.1 Self-Perception - understanding and realizing what they are doing and what they want to do.

3.2 Self-Expression - expressing the true self correctly and properly.

3.3 Stress-Management - looking at the bright side, smiling, and living happily

3.4 Decision Making - making the right decision appropriate to the present situations, and choosing the right things, not the preferred ones.

3.5 Interpersonal - having good interpersonal skills, generosity, and public mind.

In addition, the learning achievement of learners is evaluated by means of tests in the tablet computer

Part 2: Evaluation result of the developed interactive instructional model using augmented reality based on edutainment to enhance emotional quotient.

The evaluation is carried out by submitting the developed model to the ten experts for a certification on the suitability of its components, methodology, steps, activities, and for a test. The evaluation result by the expert has shows that the components have high suitability ($\overline{X}$ = 4.33, S.D. = 0.58) see table 1, the interactive instructional augmented reality steps have the highest suitability ($\overline{X}$ = 4.53, S.D. = 0.52) see table 2 same as the EDU learning activity steps ($\overline{X}$ = 4.34, S.D. = 0.49) see table 3, the emotional quotient ($\overline{X}$ = 4.27, S.D. = 0.53) see table 4, and the overall appropriateness for a test ($\overline{X}$ = 4.44, S.D. = 0.52) see table 5.

TABLE I: The Components

| Variable | Mean | SD | Level of suitability |
|---|---|---|---|
| Principles and concepts | 4.40 | 0.52 | High |
| Objectives | 4.40 | 0.70 | High |
| Process | 4.20 | 0.63 | High |
| Evaluation | 4.30 | 0.48 | High |
| Total | 4.33 | 0.58 | High |

The expert has shows high suitability ($\bar{X}$ = 4.33, S.D. = 0.58).

TABLE II: Interactive Instructional Augmented Reality

| Variable | Mean | SD | Level of suitability |
|---|---|---|---|
| Analysis environment | 4.50 | 0.53 | High |
| Preparation | 4.60 | 0.52 | Highest |
| Pre-test | 4.50 | 0.53 | High |
| Total | 4.53 | 0.52 | Highest |

The expert has shows high suitability ($\bar{X}$ = 4.53, S.D. = 0.52).

TABLE III: EDU Learning Activity

| Variable | Mean | SD | Level of suitability |
|---|---|---|---|
| Orientation | 4.40 | 0.52 | High |
| Elicitation of the prior knowledge | 4.30 | 0.48 | High |
| Turning restructuring of ideas | 4.30 | 0.48 | High |
| Application of ideas | 4.30 | 0.48 | High |
| Review | 4.40 | 0.52 | High |
| Total | 4.34 | 0.49 | High |

The expert has shows high suitability ($\bar{X}$ = 4.34, S.D. = 0.49).

TABLE IV: Emotional Quotient

| Variable | Mean | SD | Level of suitability |
|---|---|---|---|
| Behavioral observation | 4.30 | 0.48 | High |
| EQ assessment by programs in tablet computers | 4.30 | 0.48 | High |
| Evaluate learning achievement | 4.20 | 0.63 | High |
| Total | 4.27 | 0.53 | High |

The expert has shows high suitability ($\bar{X}$ = 4.27, S.D. = 0.53).

TABLE IV: Overall

| Variable | Mean | SD | Level of suitability |
|---|---|---|---|
| Instruction Model | 4.33 | 0.50 | High |
| Process and Activities | 4.56 | 0.53 | Highest |
| Actual use | 4.44 | 0.53 | High |
| Total | 4.44 | 0.52 | High |

The expert has shows high suitability ($\bar{X}$ = 4.44, S.D. = 0.52).

## 6. DISCUSSION

The research results in the following points for discussion.

6.1 The experts' evaluation demonstrates that the components, steps, and activities of the model are highly suitable. The result also aligns with the research finding of [3] suggesting that the learning management for self-development and entertaining activities consists of five key steps: 1) Orientation, 2) Elicitation of the prior knowledge, 3) Turning restructuring of ideas, 4) Application of ideas, and 5) Review.

6.2 The experts' evaluation also reveals that the model is highly suitable for the emotional quotient development. This finding corresponds to the research of [13] suggesting that the emotional quotient could contribute to learners' basic skill enhancement.

**Authors**

**Nuttakan Pakprod** is a Ph.D. candidate, Department of Information and Communication Technology for education, Faculty of Technical Education, King Mongkut's University of Technology North Bangkok (KMUTNB), Thailand. She has experience in many positions such as the Electronic Book, Computer Graphic Design, Information and Information and Communication Technology for Education research.

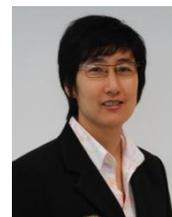

**Dr.Panita Wannapiroon** is an Assistant Professor at Division of Information and Communication Technology for Education, Faculty of Technical Education, King Mongkut's University of Technology North Bangkok (KMUTNB), Thailand. She has experience in many positions such as the Director at Innovation and Technology Management Research Center, Assistant Director of Online Learning Research Center, Assistant Director of Vocational Education Technology Research Center, and Assistant Director of Information and Communication Technology in Education Research Center. She received Burapha University Thesis Award 2002. She is a Membership of Professional Societies in ALCoB (APEC LEARNING COMMUNITY BUILDERS) THAILAND, and Association for Educational Technology of Thailand (AETT)

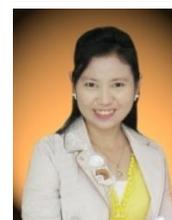